\documentclass[a4paper]{easychair} 

\usepackage{doc}
\usepackage{caption}
\usepackage{subcaption}
\usepackage{listings}
\usepackage{xcolor}
\usepackage{graphics}
\usepackage[export]{adjustbox}

\definecolor{codegreen}{rgb}{0,0.6,0}
\definecolor{codegray}{rgb}{0.5,0.5,0.5}
\definecolor{codepurple}{rgb}{0.58,0,0.82}
\definecolor{backcolour}{rgb}{0.95,0.95,0.92}

\usepackage{titlesec}
\titlespacing\section{0pt}{12pt plus 4pt minus 2pt}{4pt plus 2pt minus 2pt}
\titlespacing\subsection{0pt}{12pt plus 4pt minus 2pt}{4pt plus 2pt minus 2pt}
\titlespacing\subsubsection{0pt}{12pt plus 4pt minus 2pt}{4pt plus 2pt minus 2pt}

\newcommand{\cmmnt}[1]{}
\graphicspath{{./images/}}

\title{Deployment and configuration of MEC apps with Simu5G}
\author{
    Alessandro Noferi
\and
    Giovanni Nardini
\and
    Giovanni Stea
\and
    Antonio Virdis
}

\institute{
  Department of Information Engineering,
  University of Pisa,
  Pisa, Italy\\
  \email{a.noferi1@studenti.unipi.it},
  \email{\{name.surname\}@unipi.it}
 }

\authorrunning{Noferi et al.}

\titlerunning{Deployment and configuration of MEC apps with Simu5G}

\begin{document} 

\maketitle

\begin{abstract}
Multi-access Edge Computing (MEC) is expected to act as the enabler for the integration of 5G (and future 6G) communication technologies with cloud-computing-based capabilities at the edge of the network. This will enable low-latency and context-aware applications for users of such mobile networks. In this paper we describe the implementation of a MEC model for the Simu5G simulator and illustrate how to configure the environment to evaluate MEC applications in both simulation and real-time emulation modes.
\end{abstract}

%
%

\section{Introduction}
\label{sect:introduction}

Next-generation mobile networks will need to provide users innovative services with stringent Quality of Service (QoS) requirements, such as reduced latency and high bandwidth. To this aim, providing cloud-computing-like capabilities at the edge of the network is crucial~\cite{edgeComp}. 
Multi-access Edge Computing (MEC) is an international standard developed by the European Telecommunications Standards Institute (ETSI) that provides computing infrastructures at the edge of the network, by placing nodes, called \textit{MEC hosts}, as close as possible to end-users. Specific Application Programming Interfaces (APIs) towards the underlying network allows the MEC system to retrieve information about the status of the network and its users in real time. Such information can be used to offer new context-aware services to the end-users.  

In order to assess the performance of new MEC-based services, it is key for researchers and developers to have tools for rapid prototyping of applications in a controllable environment that includes realistic models of both the mobile network and the MEC system.

In this paper we present an ETSI-compliant implementation of the MEC system, which is included in Simu5G\footnote{https://github.com/Unipisa/Simu5G}, an OMNeT++-based simulator for the data plane of 5G networks \cite{simu5g}. We discuss our modelling choices and describe the necessary configurations for testing MEC applications  (MEC apps, hereafter) using the proposed framework. In particular, we illustrate an exemplary configuration for simulating a MEC-enabled 5G deployment, as well as the guidelines to run an emulation testbed exploiting OMNeT++/INET real-time and emulation capabilities. The latter features allow one to run MEC app prototypes, coded using any programming language, on top of a realistic 5G environment.

The rest of the paper is organized as follows: Section \ref{sect:mec} describes the main components of the ETSI MEC architecture, whereas Section \ref{sect:simu5g} provides an overview of Simu5G. Section \ref{sect:mecInSimu5G} discusses our MEC implementation within Simu5G. Section \ref{sect:deployment} illustrates the guidelines to test MEC applications within Simu5G, in both simulated and emulated mode. Section \ref{sect:conclusions} concludes the paper. 


\section{ETSI MEC Architecture}
\label{sect:mec}

\begin{figure}[ht]
  \centering
  \includegraphics[width=0.75\linewidth]{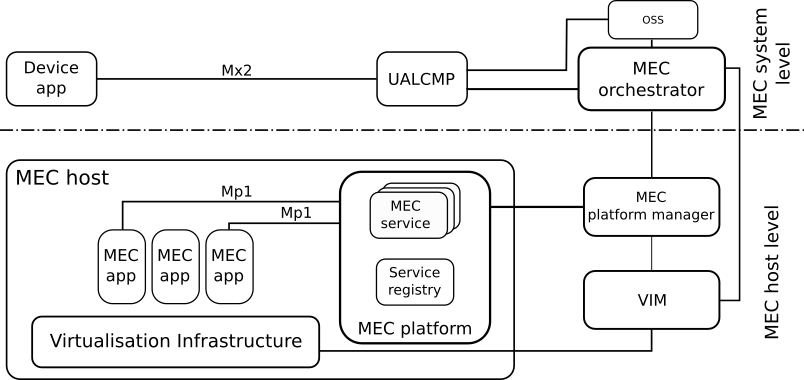}
  \captionof{figure}{MEC architecture}
  \label{fig:mecArc}
\end{figure}

According to ETSI \cite{mec003}, a MEC system is organized in two layers, namely the MEC System Level and the MEC Host Level, as shown in Figure \ref{fig:mecArc}.

The MEC system level is responsible for maintaining an overall view of the MEC hosts in the MEC system and managing the lifecycle of the MEC apps, i.e instantiation, relocation and termination. The core of the MEC system level is the MEC orchestrator, which receives requests (after a granting from the Operations Support Systems (OSS)) to instantiate or terminate applications from a Device application running on users' devices via the User Application Lifecycle Management Proxy (UALCMP). Then, the MEC orchestrator selects the MEC host where the application should be placed based on the required constraints, such as latency, available resources and required MEC services.

In the MEC host level, the MEC host contains the virtualization infrastructure where MEC apps run as virtual machines. MEC services are deployed in the MEC platform and are consumed by the MEC app through standard MEC APIs, built upon RESTful APIs \cite{mec009}. Such MEC services can be discovered by querying the Service registry, again via a RESTful API.
The MEC platform manager and the Virtualization infrastructure manager entities manage the status of the MEC host and are queried by the MEC orchestrator to retrieve information (e.g. available resources, MEC services, etc.) during the MEC app lifecycle operations.

Communications among MEC system entities occur via standardized reference points, like Mp1 between MEC app and the MEC platform, and Mx2 between Device app and UALCMP.

Within a 5G network, MEC hosts can be deployed at different location in a flexible way, e.g. at the base station --- or an aggregation of them--- or in the core network \cite{mecIn5g}. 


\section{Overview of Simu5G}
\label{sect:simu5g}
This section provides a brief description of the main components of Simu5G, with special focus on the modules involved in the MEC architecture.

\begin{figure}[!ht]
\centering
\begin{minipage}[t]{.5\textwidth}
  \centering
  \includegraphics[width=0.9\textwidth, valign=b]{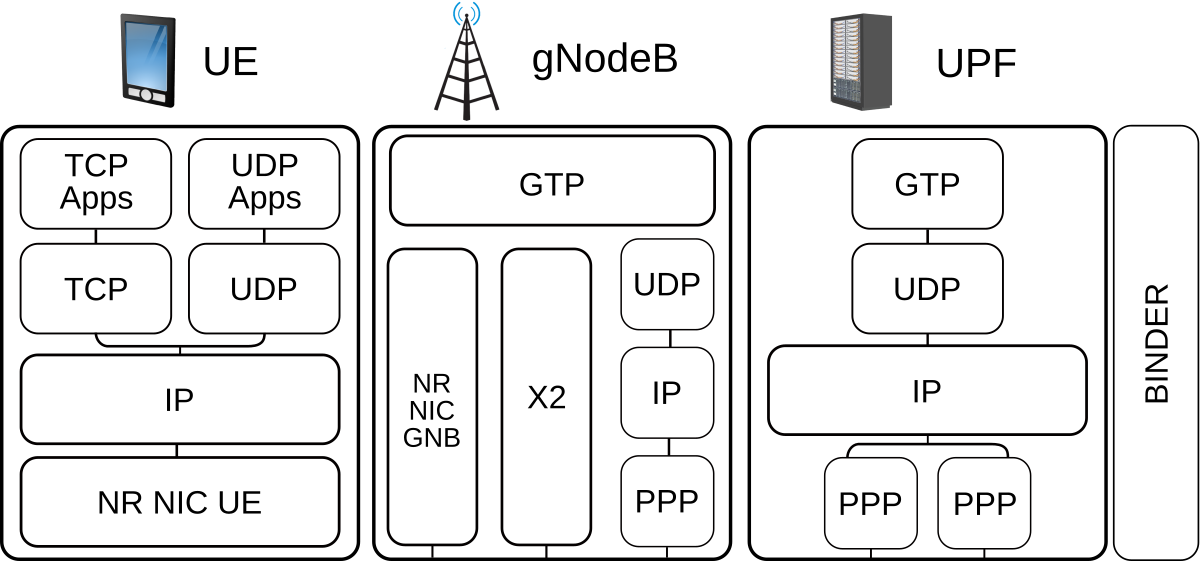}
  \captionof{figure}{Simu5G main modules}
  \label{fig:simu5G}
\end{minipage}%
\begin{minipage}[t]{.5\textwidth}
  \centering
  \includegraphics[width=0.7\textwidth, valign=b]{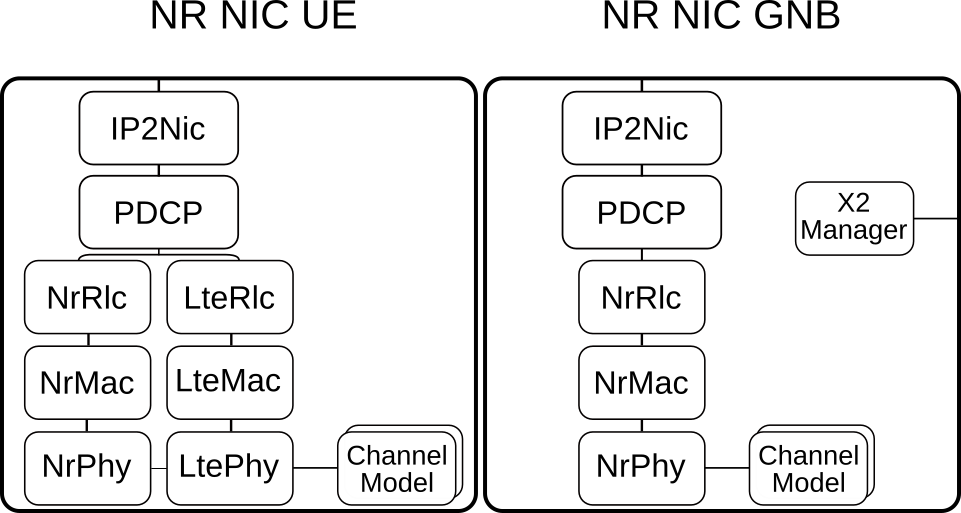}
  \captionof{figure}{NIC modules}
  \label{fig:nic}
\end{minipage}
\end{figure}

Simu5G \cite{simu5g} is the evolution of the popular SimuLTE 4G network simulator \cite{simulte-book2019} that incorporates 5G New Radio access. It models the user plane of both the Core Network (CN) and the Radio Access Network (RAN). As depicted in Figure \ref{fig:simu5G}, within Simu5G a 5G network is composed of User Equipments (UEs), gNodeBs (gNBs) and User Plane Functions (UPFs) modules. The latter implements the GPRS tunneling protocol (GTP) enabling the routing of packets from the RAN to Internet through the CN.

UEs and gNBs communicate via the layer-2 New Radio (NR) protocol stack, implemented along with the physical layer in the \textit{NrNic} module. The internal design of the NIC for both the gNB and UE is depicted in Figure \ref{fig:nic} and it is composed of one submodule for each layer of the NR protocol stack. On the UE side, the NIC also contains submodules for the LTE protocol stack, so as to allow both mixed 4G/5G and E-UTRA/NR Dual Connectivity scenarios.

The topmost part of the TCP/IP stack, i.e. from the application layer to the IP included, is provided by the INET library, and so are the UE mobility models. One can also use other mobility libraries available for OMNeT++, such as Veins\cite{veins} or Artery \cite{artery} for vehicular mobility.
gNBs communicate with each other via the X2 interface, e.g. for handover and interference-coordination mechanisms.

Control-plane functions and protocols are generally not modeled. Instead, a special module, called \textit{Binder} takes care of maintaining network-wide data structures (e.g., which UE is associated with which gNB, etc.). Network nodes (e.g., UEs and gNBs) can query the Binder via direct method calls, thus mimicking control-plane functions without the need to setup complex protocol state machines. 

Simu5G can also be run as a real-time 5G network emulator by leveraging OMNeT++'s real-time event scheduler and INET's external interfaces. This enables an external application to exchange packets with a module in an emulated 5G network, or to use the 5G network as a transport network between two external applications.

\section{Modelling MEC within Simu5G}
\label{sect:mecInSimu5G}
The proposed MEC model aims to provide MEC developers a tool for rapid prototyping of MEC apps, which can exploit MEC services made available by Simu5G. Application endpoints (e.g. Mp1 and Mx2 interfaces) and MEC service communication have been implemented according to the ETSI specifications \cite{mec011,mec016,mec009}. Since Simu5G can run as a real-time network emulator, one can also interface \texttt{real} MEC apps with it. This way, one can test MEC apps in a computation/communication framework that includes 5G transport, a MEC infrastructure, and the MEC services provided by the former to the latter.

In this section, we describe how MEC entities have been modelled.

\subsection{MEC system-level modelling}
\label{subsect:mecSystem}
\begin{figure}[!ht]
\centering
\begin{minipage}[t]{.5\textwidth}
  \centering
  \includegraphics[width=0.9\textwidth,valign=b]{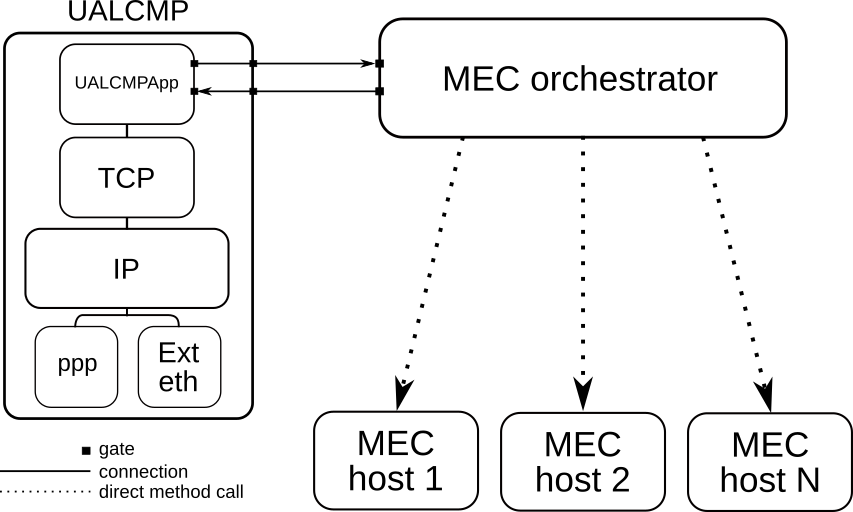}
  \captionof{figure}{MEC system-level modeling}
  \label{fig:mecSys}
\end{minipage}%
\begin{minipage}[t]{.5\textwidth}
  \centering
  \includegraphics[width=0.8\linewidth,valign=b]{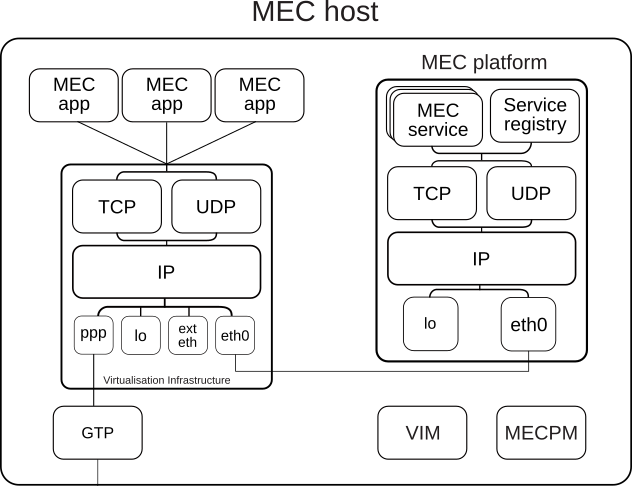}
  \captionof{figure}{MEC host-level modeling}
  \label{fig:mecHost}
\end{minipage}
\end{figure}

Figure \ref{fig:mecSys} shows our models for the UALCMP and the MEC orchestrator. The UALCMP is a compound module including the TCP/IP stack and one application module that  provides the RESTful API consumed by the Device app (either internal or external to the simulator, through an INET external interface) to trigger instantiation and termination of MEC apps via the Mx2 reference point. The MEC orchestrator is a simple module connected to the UALCMP via gates. In scenarios with multiple MEC systems, the MEC orchestrator manages a subset of the MEC hosts in the simulation, hence it must specify such subset using the \texttt{mecHostsList} parameter, which is a space-separated string specifying the names of the MEC hosts under its control. 
The UALCMP and the MEC orchestrator communicate via OMNeT++ messages, whereas interactions with MEC host-level entities are realized via direct method calls (see Figure \ref{fig:mecSys}). Such interfaces are not fully compliant with the standard. On one hand, the reason is that they have not been completely standardized by ETSI at the time of writing. On the other hand, this allows us to model all the functionalities required by the application endpoints without overly complicating the system.

Upon receiving a MEC app instantiation request from the UALCMP (which in turn receives the request by a Device app), the MEC orchestrator selects the most suitable MEC host among those associated with its MEC system, according to the MEC app requirements. The latter are specified in the so-called \textit{AppDescriptor} JSON file and can include a list of required MEC services and/or computing resources (RAM, CPU and disk), which are checked against the available resources of the MEC hosts. The location in the file system of the AppDescriptor file can be specified by either the Device app triggering the MEC app instantiation or the MEC orchestrator directly. 
eAs per ETSI specifications in \cite{mec010}, the \textit{AppDescriptor} file for our implementation must specify at least the following fields: \textit{appDid}, \textit{appName}, \textit{appProvider}, \textit{virtualComputeDescriptor} and the\textit{appServiceRequired}, as shown in Figure \ref{fig:appD}.

Once the MEC host has been identified, the MEC orchestrator triggers the MEC app instantiation, which is accomplished by dynamically creating the OMNeT++ module running the application logic (see \ref{subsect:mecHostLevel}) and marking the required MEC host resources as allocated. Similarly, upon a MEC app termination request, the MEC orchestrator informs the MEC platform manager to delete the module and to release the resources previously allocated.
We model the processing delay at the MEC orchestrator by scheduling a timer with configurable duration: when the latter expires, the MEC orchestrator replies to the UALCMP by acknowledging the instantiation/termination of the MEC app.

The Device app is usually embedded in the UE client. However, its functions are quite standard, i.e., managing instantiation/termination of MEC apps by interfacing with the UALCMP. For this reason, our MEC architecture includes a basic Device app able to query the UALCMP via the RESTful API and communicating with the UE app over an UDP socket by means of a simple interface including messages for creation, termination and related acknowledgments regarding a MEC app. This way, a user only needs to code the data-plane colloquium between the UE app and the MEC app, as well as their logic.

\subsection{MEC host-level modelling}
\label{subsect:mecHostLevel}
The MEC host module models the MEC host-level of the MEC architecture, as shown in Figure \ref{fig:mecHost}. It is configurable (via NED/INI files) with a maximum amount of resources, i.e. RAM, storage and CPU, that can be allocated to MEC apps. Its main components are the Virtualisation Infrastructure Manager (VIM) and the MEC platform modules. As previously mentioned, communications with the MEC platform can also involve real MEC apps when Simu5G runs as an emulator. For this reason, all the entities, i.e. MEC service and Service Registry, implement a RESTful HTTP server as a TCP application running on top of the TCP/IP protocol stack of the MEC platform. In particular, the framework provides a module called \textit{MecServiceBase} and it implements all the non-functional requirements needed for running an HTTP server (e.g. queueing, TCP connection management, gNB connections etc.). This allows the user to rapidly deploy an ETSI-compliant MEC service by only implementing the HTTP methods (e.g. GET, POST) according to the service behavior.

Two ETSI MEC services are currently implemented: \textit{Radio Network Information service (RNIS)} \cite{RNI}, that allows users to gather up-to-date information regarding radio network condition and the UEs connected to the base station associated to the MEC host, and the \textit{Location service} \cite{locAPI}, that provides accurate information about UEs and/or base station locations, enabling active device location tracking and location-based service recommendations.

The \textit{ServiceRegistry} module implements the REST resources -- and HTTP methods -- needed to allow MEC apps to discover MEC services via the Mp1 reference point \cite{mec011}.

MEC apps are deployed as applications over TCP and UDP transport-layer protocols of the Virtualization Infrastructure compound module. They implement the \textit{IMECApp} interface and are dynamically created by the VIM upon instantiation requests triggered by the MEC orchestrator through the MEC platform manager. The VIM manages MEC host resources and keeps track of the MEC apps currently running on the virtualization infrastructure through a data structure called \textit{mecApps}. For each MEC app, the latter contains a \textit{mecAppEntry} structure with the pointer and the associated gate indexes of the module,\cmmnt{to be subsequently} used for the deallocation phase, the endpoint of the deployed MEC app\cmmnt{(in \texttt{IP:port} format)} returned to the UE in order to communicate with it, and the allocated resources. Such resources are used by the VIM to compute the delay used to model processing time. More in detail, the CPU constraint is expressed in terms of instructions per second, so that we can model the processing time of a MEC app, i.e. the time required by the MEC app to execute a set of instructions. This is done by the VIM according to two different paradigms: \textit{Segregation}, where the MEC app obtains exactly the amount of computing resources it has configured in the  \textit{AppDescriptor} file, even when no other MEC apps are running concurrently and 
 \textit{Fair sharing}, in which MEC apps share all the available computing resources proportionally to their requested rate, possibly obtaining more capacity than the amount stipulated in the \textit{AppDescriptor} file.

Finally, a MEC host also has a GTP module that allows it to be placed anywhere in the CN of the 5G network. This is useful, for example, to test different MEC hosts deployments.

\section{Deployment scenarios}
\label{sect:deployment}
As already mentioned, Simu5G can run in both simulated and emulated mode. So, the user can develop and test MEC-related applications in either mode, according to her needs. For instance, it could be the case that only a real MEC app should be tested, leaving the UE application as a stub method inside the simulator, possibly because its behavior is not needed or it is not yet available. This section illustrates the configurations required to run some MEC scenarios in both Simu5G execution modes.
\subsection{Running a fully simulated MEC system}
\label{subsect:fullySim}
\begin{figure}
\centering
\begin{minipage}{.5\textwidth}
  \centering
  \includegraphics[width=0.9\linewidth]{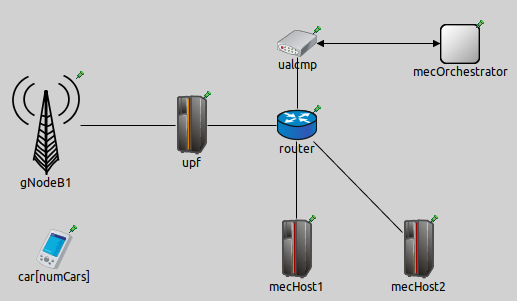}
  \captionof{figure}{Simu5G network}
  \label{fig:network}
\end{minipage}%
\begin{minipage}{.5\textwidth}
   \centering
    \frame{\includegraphics[width=0.85\linewidth]{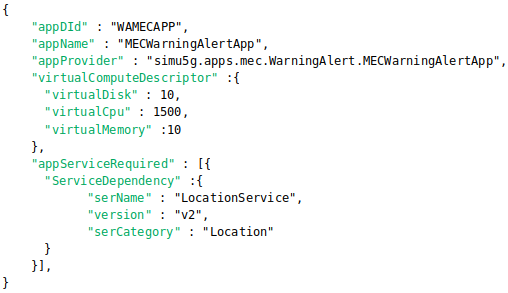}}
  \captionof{figure}{AppDescriptor for WarningAlertApp}
  \label{fig:appD}
\end{minipage}
\end{figure}

The network used for the following simulation is depicted in Figure \ref{fig:network}. The MEC system contains a UALCMP, a MEC orchestrator and two MEC hosts, namely \textit{mecHost1} and \textit{mecHost2}, that are associated to a gNB.
\textit{Car} is a vector of UEs, each of them running an application called \textit{UEWarningAlertApp}. The latter instantiates a MEC app named \textit{MECWarningAlertApp}, which is supposed to be onboarded in the MEC system during network initialization and to send alert messages to cars when they enter a danger zone. Cars also run the Device app responsible to request the MEC app instantiation to the UALCMP on behalf the UEs. Besides the module implementing the logic of the \textit{MECWarningAlertApp}, the \textit{AppDescriptor} is also required. In accordance with \ref{subsect:mecSystem}, the \textit{AppDescriptor} file that describes the \textit{MECWarningAlertApp} is shown in Figure \ref{fig:appD}.

\begin{figure}[!ht]
\centering
\begin{minipage}[t]{.5\textwidth}
    \centering
    \frame{\includegraphics[width=0.95\linewidth,valign=t]{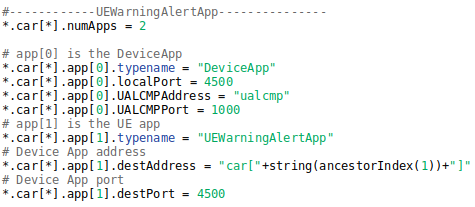}}
    \caption{car configuration}
    \label{fig:carConf}
\end{minipage}%
\begin{minipage}[t]{.5\textwidth}
    \centering
    \frame{\includegraphics[width=0.95\linewidth,valign=t]{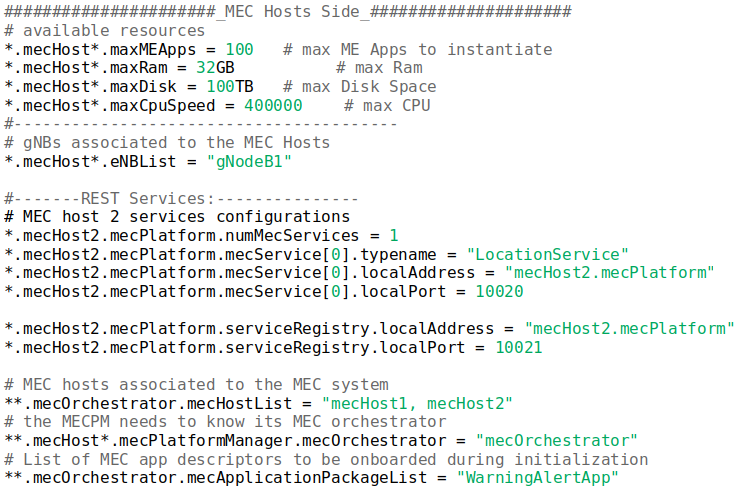}}
    \caption{MEC system configuration}
    \label{fig:mecConf}
\end{minipage}
\end{figure}

The MEC app requires 10 MB of RAM, 10 MB of storage, 1500 instructions per second of CPU and it consumes the Location Service. Next, it is necessary to configure both the car[*] modules and the MEC entities. For the former, the snippet of the INI file in Figure \ref{fig:carConf} shows the configuration relevant to the MEC system.
Both the MEC hosts have the same computational capacity, but only mecHost2 has a Location Service running on its MEC platform, hence it will be the one chosen by the MEC orchestrator to deploy the MEC app.

If the user wanted to request the onboarding of the \textit{appDescriptor} file from the Device app, instead of it being loaded at initialization time, it has to insert the path to the file in the \textit{appPackageSource} parameter of the Device app and remove the \textit{appDescriptor} file name from the list of the MEC orchestrator \textit{mecApplicationPackageList} parameter.
The complete scenario configuration can be found in the \textit{simulations/NR/mec} folder on the GitHub\footnote{https://github.com/Unipisa/Simu5G/tree/master/simulations/NR/mec/multiMecHost} repository.
\subsection{Running a MEC system in emulation mode}
\label{subsect:emuMec}
\begin{figure}[ht]
  \centering
  \includegraphics[width=0.75\linewidth]{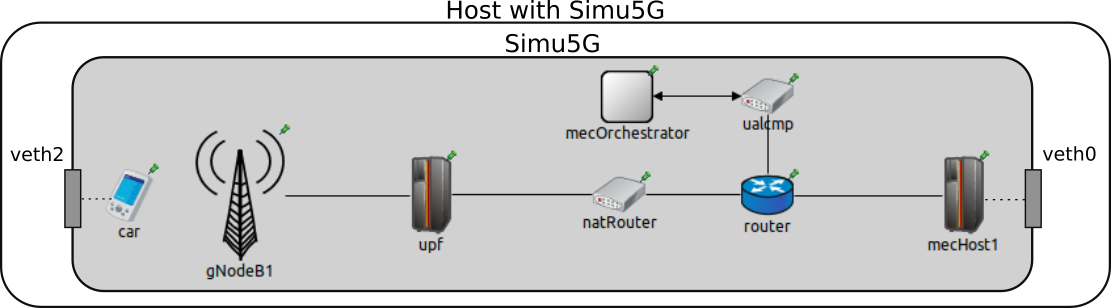}
  \captionof{figure}{Simu5G network for emulation}
  \label{fig:vethEmu}
\end{figure}

We now describe the configuration required to run a MEC system in emulation mode within Simu5G. More in detail, the following operations will be highlighted: routing rules, in both the simulator and the host running the real applications, and the configuration to instruct the MEC orchestrator to manage the instantiation of a real MEC app. 
In our test, both the UE app and the MEC app are real applications running on the same host where Simu5G runs. The real applications replicate the behavior described in section \ref{subsect:fullySim} in a network scenario including one car and one MEC host, as shown in Figure \ref{fig:vethEmu}. The Device app runs inside the UE and communicates with the real UE app through the serialization and deserialization of the messages of the interface mentioned in Section \ref{subsect:mecSystem}. Although we focus on a single-host testbed, it is also possible to run the applications on different hosts by adjusting routing configurations accordingly.
\vspace{-0.3cm}

\begin{figure}[ht]
  \centering
  \frame{\includegraphics[width=0.75\linewidth]{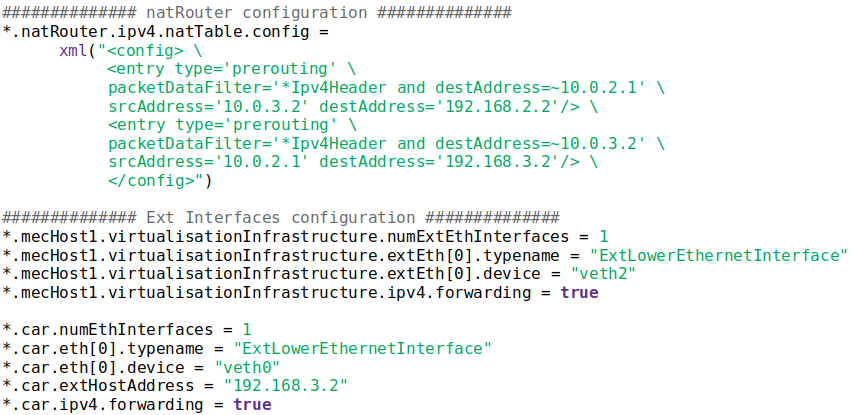}}
  \captionof{figure}{natRouter and \textit{ExtLowerEthernetInterface} configuration}
  \label{fig:natExtIntConf}
\end{figure}

The interaction with the real world occurs via two INET's \textit{ExtLowerEthernetInterface} modules included into the car and mecHost modules. Such interfaces can receive real packets by network interface cards attached to them. In our case, such network interfaces are created as Virtual Ethernets (veth). Data packets directed to the simulator are routed to the relative veth attached to the \textit{ExtLowerEthernetInterface} modules. Since all the applications run in the same host, a \textit{natRouter} module must be used to bypass the host operating system and steer the traffic towards the simulator. This way, both real applications send packet to the IP addresses of the \textit{natRouter}, which in turn performs Network Address Translation by changing the destination addresses to the proper real application's addresses. The INI configuration for the \textit{natRouter} router is shown in the topmost part of Figure \ref{fig:natExtIntConf}. In the latter, with reference to Figure \ref{fig:vethEmu}, the IP addresses \texttt{10.0.2.1} and \texttt{10.0.3.2} are the addresses of the left and right \textit{natRouter} interfaces, respectively. 

To conclude the Simu5G network setup, the configuration of the routing tables of all the network devices in the simulated network is needed to enable the forwarding for packets destined to the real applications. Such files, with all the other configuration of this scenario, can be found in the \textit{emulation/mec} folder on the GitHub\footnote{https://github.com/Unipisa/Simu5G/tree/master/emulation/mec/extUeAppMecApp} repository. 

Finally, the bottom part of Figure \ref{fig:natExtIntConf} depicts the INI configuration of the \textit{ExtLowerEthernetInterface} modules to allow the communication with the real world.

As far as the MEC system is concerned, to allow the instantation of a real MEC app in the \textit{appDescriptor} file related to a real MEC app,  a field called \textit{emulatedMecApplication} must be inserted. The latter contains \textit{ipAddress} and \textit{port} sub-fields identifying the real MEC app endpoint. 
This way, the MEC orchestrator is made aware that the MEC app to instantiate is running outside Simu5G and it does not need to request the creation of the MEC app module inside the simulator. Then, the MEC orchestrator communicates to the Device app the IP-port pair the UE app will need to use to communicate with the MEC app, i.e. the address of the natRouter interface in this case, although real hosts' addresses are also allowed.

\begin{figure}[ht]
  \centering
  \frame{\includegraphics[width=0.75\linewidth]{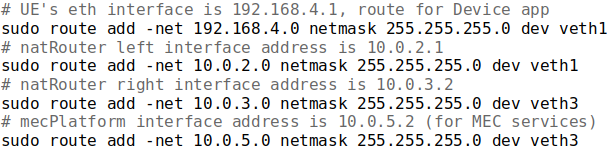}}
  \captionof{figure}{\textit{veth} interfaces configuration}
  \label{fig:routes}
\end{figure}

Once the Simu5G environment is configured, the OS of the host running all the applications must be configured too. The following commands refers to a host equipped with Linux Ubuntu 18.04 OS. \textit{veth} interfaces are created through the command: \texttt{ip link add veth0 type veth peer name veth1} (the same for the couple \textit{veth2}-\textit{veth3}). After the interfaces have been created, we assign an IP address to them and enable them by \texttt{ip addr add 192.168.3.2 dev veth1} (and 192.168.2.2 for \textit{veth3}) and \texttt{ip link set veth0} up (for all the 4 interfaces), respectively. Finally, routes to forward the packets within the simulator and its modules have to be added. In particular, packets must reach the Device app, the natRouter and the MEC platform modules, as shown in Figure \ref{fig:routes}.
\vspace{-1cm}



\section{Conclusions}
\label{sect:conclusions}
In this paper, we described the modeling of a MEC framework to be integrated with the Simu5G simulator, providing a complete tool for rapid prototyping of MEC applications in MEC systems running on a 5G network. The implementation of standard ETSI APIs, combined with the ability to run Simu5G in emulation mode, allows MEC developers to test and evaluate real MEC apps running on real hosts -- possibly interfacing with external frameworks like Intel OpenNESS. Such real applications exploit Simu5G to emulate the 5G transport network between them and to consume MEC services for obtaining information related to the 5G network. 
We also presented the configurations needed to run an MEC system --- and MEC-related applications --- with Simu5G, in both simulation and emulation modes. 

\label{sect:bib}
\bibliographystyle{plain}
\bibliography{refs}



\end{document}